# PANDA: Extreme Scale Parallel K-Nearest Neighbor on Distributed Architectures

Md. Mostofa Ali Patwary[1,†], Nadathur Rajagopalan Satish[1],
Narayanan Sundaram[1], Jialin Liu[2], Peter Sadowski[3], Evan Racah[2],
Suren Byna[2], Craig Tull[2], Wahid Bhimji[2], Prabhat[2], Pradeep Dubey[1]

[1]Intel Corporation, [2]Lawrence Berkeley National Laboratory, [3]UC Irvine
[†]Corresponding author: mostofa.ali.patwary@intel.com

*Abstract*—Computing $k$-Nearest Neighbors (KNN) is one of the core kernels used in many machine learning, data mining and scientific computing applications. Although kd-tree based $O(\log n)$ algorithms have been proposed for computing KNN, due to its inherent sequentiality, linear algorithms are being used in practice. This limits the applicability of such methods to millions of data points, with limited scalability for Big Data analytics challenges in the scientific domain. In this paper, we present parallel and highly optimized kd-tree based KNN algorithms (both construction and querying) suitable for distributed architectures. Our algorithm includes novel approaches for pruning search space and improving load balancing and partitioning among nodes and threads. Using TB-sized datasets from three science applications: astrophysics, plasma physics, and particle physics, we show that our implementation can construct kd-tree of 189 billion particles in 48 seconds on utilizing ∼50,000 cores. We also demonstrate computation of KNN of 19 billion queries in 12 seconds. We demonstrate almost linear speedup both for shared and distributed memory computers. Our algorithms outperforms earlier implementations by more than order of magnitude; thereby radically improving the applicability of our implementation to state-of-the-art Big Data analytics problems. In addition, we showcase performance and scalability on the recently released Intel® Xeon Phi™ processor showing that our algorithm scales well even on massively parallel architectures.

*Keywords*-Big Data Analytics, KNN, kd-tree, Classification, Parallel Algorithms, and Load Balancing.

## I. INTRODUCTION

The $k$-nearest neighbor (KNN) algorithm is a fundamental classification and regression method for machine learning. KNN is used for many tasks such as text classification [1], prediction of economic events [2], medical diagnosis [3], object classification in images, prediction of protein interactions, and so on. KNN works by finding the $k$ nearest points to a given query point in the feature space. The continuous or discrete predicted output value for the query is then computed using the corresponding values of the neighbors (e.g. using a majority vote for discrete classification problems, or an average of values in a continuous regression setting).

With the advent of sophisticated data collection mechanisms, machine learning on large datasets has becoming very important. Scientific disciplines such as cosmology and plasma physics perform simulations in the range of billions of particles and produce many terabytes of data per time step [4]. Modern Particle Physics experiments such as Daya Bay and the Large Hadron collider deploy tens of thousands of sensors that are capable of capturing data at nano-second precision; such experiments can produce 1000s of TBs per year of data. For such large datasets, the data is typically distributed in the memory of multiple nodes in a cluster of machines, and machine learning methods such as KNN must in turn be able to leverage this data and perform distributed computations on them. Moreover, in the case of simulation data, each simulation timestep must run in a small amount of time in the range of minutes to be useful to a domain scientist, and this imposes strict runtime restrictions on the machine learning model as well. At the scale of billions of points, it is important to leverage all the available parallelism in the hardware – at the cluster level and inside each compute node – in order to provide results in a suitable timeframe.

Since KNN is a fundamental algorithm used in a variety of machine learning contexts, a significant amount of time has gone into making it run fast on parallel hardware [5], [6], [7]. However, most of this work has been done to reduce algorithmic complexity or parallelization in a shared memory setting. An important algorithmic contribution has been the introduction of acceleration data structures such as kd-trees [8] (which hierarchically partitions points in a $k$ dimensional space) to reduce the order complexity of near neighbor searches per query from linear to logarithmic in the number of points. Such data structures work well in a variety of scientific applications where the dimensionality of data is not very high, and are critical for large data sets where we cannot afford to scan the entire data set for each query. Furthermore, parallel algorithms that can utilize multiple cores of a single shared memory machine to speed up kd-tree based KNN have been developed [5], [6], [7]. However, there has not been much work on parallelizing this in a distributed setting. Indeed, previous work for distributed KNN has mainly focused on brute force linear-time approaches per query without using acceleration data structures [9], [10]. A recent work [11], a clustering algorithm, performs radius search based nearest neighbor computation in a distributed setting. However, in a distributed setting, radius based search is an easier problem; the fixed radius allows easy pruning of the points that need to be searched. [11] relies on the radius being small; indeed, for large radius values, the entire dataset may need to be gathered to one node. This makes it unsuitable for KNN, where the radius is not known apriori.

There are many challenges in developing a distributed algorithm for kd-tree based KNN computations. Consider the straightforward data-parallel approach where the data is evenly distributed among the nodes; and each node constructs its own local kd-tree. While this makes kd-tree construction trivially parallel, each nearest neighbor query must then be run on all nodes and a top-$k$ algorithm must then be run on the results of each node to find the global nearest neighbors. This results in unnecessary network traffic on the interconnect between the nodes as well as unnecessary kd-tree traversal steps - indeed to find the $k$ nearest neighbors, it is easy to see that we compute and transfer $P*k$ near neighbors and throw away all but the nearest $k$. We can do much better if we can spatially partition the data and distribute the partitions among the nodes. In fact, we can build a global kd-tree containing all data points from these spatial partitions. In this case, on average, we will only query $O(\log(P))$ nodes for each query, which improves scalability. This also reduces the communication time proportionately. The tradeoff here is that more time is spent in construction of the kd-tree itself, but the kd-tree in most application contexts is reused heavily, making this worthwhile.

In this paper, we demonstrate a fully distributed implementation for kd-tree based KNN computations. This is, to the best of our knowledge, the first such work. In addition to adopting the global kd-tree approach, we also focus on various optimizations to utilize all levels of parallelism, both at the cluster and intra-node level to make both construction and querying fast. Additionally, we improve network performance by overlapping computation and communication through software pipelining. Our resulting implementation is more than an order of magnitude faster than previous approaches on a single node; and scales near linearly with number of cores and nodes. Using ∼50,000 cores, we can construct kd-tree of 189B points (∼3 TB dataset) in 48 seconds and run 19B queries on that dataset in ∼12 seconds.

The main technical contributions are as follows:

- This is the first distributed kd-tree based KNN code that is demonstrated to scale up to ∼50,000 cores.
- This is the first KNN algorithm that has been run on massive dataset ranging up 100B+ dataset from diverse scientific disciplines.
- We successfully demonstrate both strong and weak scalability of KNN at this scale.
- Our implementation is more than an order of magnitude faster than state-of-the-art KNN implementation.

## II. SCIENCE MOTIVATION

The construction and application of kd-trees are highly dependent on the distribution of the underlying dataset. Instead of choosing a random dataset, we have decided to utilize scientific datasets from HPC simulations (cosmology and plasma physics) and experiments (Daya Bay). Cosmology and Plasma physics datasets are representative of the spatial distribution of particles subject to gravitational and electromagnetic forces. The Daya Bay dataset is representative of discrete spatio-temporal events.

**Cosmology**: Cosmology has been at the frontier of physical sciences over the past two decades. Cosmology observations and simulations produce and analyze massive amounts of data. The quest for ever-fainter Cosmic Microwave Background (CMB) signals from the ground, balloons, and space has driven an exponential growth in data volumes in the last 25 years. Projects such as Sloan Digital Sky Survey (SDSS) collect terabytes of data related to observations of billions of objects. Analysis of these large datasets lead to an understanding of cosmology mysteries such as dark matter and the accelerated expansion of the universe.

Gravitational instability, where small initial density fluctuations are rapidly enhanced in time, drive the formation of structure in the universe. The resultant density field in the structure contains large void spaces, many filaments, and dense clumps of matter within filaments. The existence of those localized, highly over-dense clumps of dark matter, termed *halos*, is an interesting astronomical object in current cosmological models. Dark matter halos occupy a central place in the paradigm of structure formation: gas condensation, resultant star formation, and eventual galaxy formation all occur within halos. A basic analysis task is to find and classify these clusters of particles, and measure their properties like mass, velocity dispersion, density profile, and others. Since cosmology datasets surveying or simulating the universe contain billions of low dimensional particles (described by spatial locations and velocities), highly scalable KNN algorithms for classification are required.

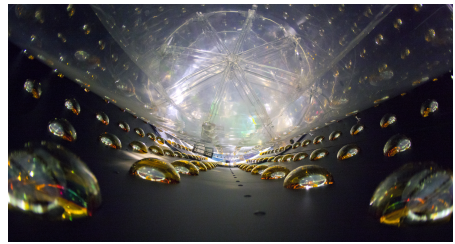

Figure 1. The interior of one of the cylindrical Daya Bay antineutrino detectors (Credit: Roy Kaltschmidt, Lawrence Berkeley National Laboratory)

**Plasma Physics**: Magnetic reconnection is a mechanism that releases magnetic energy explosively as field lines break and reconnect in plasmas. This fundamental process plays a significant role in the dynamics of various systems including laboratory fusion experiments, the Earth's magnetosphere reactions to solar eruptions, and the solar corona. Computational plasma physicists are often interested in understanding the behavior of highly energetic particles near the magnetic reconnection that could lead to understanding the causes of the phenomenon. Plasma physics simulations, such as VPIC, simulate trillions of particles running on large-scale supercomputers and stores electron properties. The capability to classify particle features such as flux ropes, and high density clusters in phase space is an important analytics task that can facilitate storage of spatial subsets, and Adaptive Mesh Refinement strategies. The primary challenge in accomplishing this task is the development of highly scalable classification and regression algorithms that could scale to billions of low dimensional particles (spatial locations, energy and velocities).

**Particle Physics**: Particle physics explores the nature of fundamental sub-atomic particles. Modern particle physics experiments involve the deployment of instruments with many channels, recording data at high-frequency. These experiments produce petabyte scale datasets. The signals of interest in these huge datasets are often rare new particles or infrequent interactions and the infeasibility of storing all the data recorded by these experiments make fast and accurate classification critical. For this study we use data from the Daya Bay Reactor Neutrino Experiment (Figure 1) which is designed to observe and characterize neutrino oscillations, a phenomena that goes beyond the so-called 'Standard Model' of Particle Physics.

## III. ALGORITHMS

Given the large size of the datasets that are regularly generated or observed in scientific applications, it is essential to utilize all available levels of parallelism in a system to achieve best performance. In many large scientific simulations involving billions of particles, $k$-nearest neighbor queries need to be run for a large percentage of the particles at each simulation step, imposing high performance requirements for distributed KNN queries. Since such simulations typically deal with low dimensional data, acceleration structures such as kd-trees work well and are essential to reduce the order complexity of KNN queries.

Most previous work on distributed KNN querying has used exhaustive search over all particles rather than using kd-trees [9], [10]. While kd-trees offer lower order complexity, constructing and querying such trees does involve more global data redistribution and control divergence. It is then imperative to perform careful optimizations for both kd-tree construction and querying in order to reap the benefits of the improved order complexity. Further, there are tradeoffs to be made between kd-tree construction times versus the quality of trees produced (which dictates query times). In typical simulation scenarios, the particles move at the end of each iteration, and one would like to reconstruct a new kd-tree every few iterations to keep queries fast. This means that while it is important to optimize the performance of both kd-tree construction and queries in a distributed setting, we have some flexibility for construction that we can exploit.

In this section, we describe our algorithm (called PANDA) for distributed kd-tree construction and querying for k-nearest neighbors. Our algorithm take advantage of multiple nodes, multiple cores and SIMD. We also describe the trade-offs we made between kd-tree construction and querying.

### A. Distributed kd-tree construction

As mentioned earlier, ours is the first fully distributed and optimized kd-tree construction algorithm scaling up to $\sim 50,000$ cores. We achieve this by taking advantage of parallelism at all levels in the hardware: multinode, multicore and SIMD. The kd-tree has a global component and local components. The global component (called *global kd-tree*), refer to the partition tree of the dataset among the nodes according to their geometry. The local components (called *local kd-tree*) refers to the kd-tree formed from among the points assigned to a given node.

There are two strategies to achieve a distributed kd-tree - (1) One option is to perform no global redistribution of points and let each node construct its own independent kd-tree of a (load-balanced) subset of points. This achieves good performance on tree construction, however querying is much slower as each query has to be sent to all the nodes to get the right answer. (2) The other option is to create one large kd-tree for all the points. This necessarily involves global redistribution of points, however this cost is more than balanced out by the reduced runtime of querying as each query only needs to be answered by a small subset of the nodes. We follow the latter approach since we have some leeway in construction costs as described previously.

Figure 2 refers to the overall structure of distributed kd-tree construction. This phase consists of 4 main steps:

**i) Global kd-tree construction**: We assume that each node reads in an approximately equal number of points (in no particular order). At every level of the kd-tree, the data points need to be split approximately equally into two subsets. We perform this split by choosing a dimension and a split point along that dimension. The algorithms used for these choices are presented below in detail (Section III-A1). Once the split point is chosen, nodes need to redistribute points so that every node only has points belonging to one of the subsets. This necessarily involves communication across the network. The process continues until each node has a non-overlapping subset of points.

**ii) Local kd-tree (data parallel)**: We then continue this process of calculating split points and shuffling in each node on its local points. Within a shared memory environment, the shuffling stage only involves moving the index, not the points themselves. At the top levels of the local kd-tree, there are not enough branches to exploit thread level parallelism, hence we use data parallelism to calculate the split & shuffle points and proceed in breadth-first fashion (one level at a time). Once there are sufficient branches (typically, number of threads $\times 10$), then we move to the next stage.

**iii) Local kd-tree (thread parallel)**: In this stage, each thread proceeds to create the kd-tree from a distinct, non-overlapping set of points. In order to ensure cache locality, this tree construction proceeds in a depth-first fashion. We declare a node to be a leaf node when the number of points in it reaches a threshold (maximum bucket size).

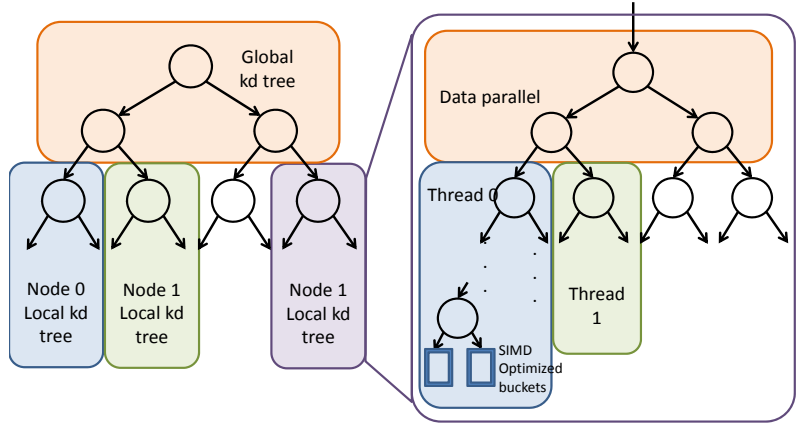

Figure 2. (a) Distributed kd-tree. The top levels of the tree are global, whereas the bottom levels are localized to different nodes. (b) Local kd-tree. The top levels are constructed through data parallelism, whereas the lower levels are optimized using thread-level parallelism and SIMD.

*iv) SIMD packing*: Once the points in each bucket are fixed, we shuffle the dataset so that points within a bucket are localized in memory. This improves the query performance as we need to perform exhaustive distance computation with all points in a bucket (if selected) at query time.

*1) Algorithmic choices:* We discuss some of the algorithmic choices we used in distributed kd-tree construction.

**Choice of split dimension**: As mentioned earlier, at every level of the kd-tree the data needs to be split approximately equally into two subsets. We perform this split by choosing a dimension and a split point along that dimension. The former choice may be based on maximum range (e.g. ANN[12]) or a more nuanced metric allowing for better partitioning. We use the dimension with maximum variance as the best dimension to split in the kd-tree. As this computation could be expensive, we take a subset of points to compute variances. This is similar to the strategy used in FLANN[13]. Although this adds up to 18% to the kd-tree construction time, we observe that this can improve query performance by up to 43% (e.g. particle physics dataset).

**Choice of split point**: We then have to choose the split point along that dimension. The ideal point is the median along that dimension that promises to divide the dataset into 2 equal subsets. Calculating the median is expensive, hence we use heuristics to approximate this. We utilize a sampling heuristic to estimate the data distribution along a dimension and choose a point close to the median as the split point. Our heuristic is similar to that of [11]. For the global kd-tree, every node samples a small set of points ($m$ points each; $m = 256$ for global kd-tree) and sends it to all the other nodes. Given the set of $P \times m$ points, each node constructs the histogram along the chosen dimension using these as the (non-uniform) interval points. This histogram information is broadcast to all nodes, thereby allowing all the nodes to construct a global histogram of all points in the dataset at non-uniform bins. We then choose the approximate median using the global distribution (interval point closest to 50%). This technique is also applied at the local level to obtain approximate median at the node level (1024 samples for local kd-tree). The threads then co-operatively build the histogram for the local points using these interval points to obtain an approximate median. We further optimize the operation of finding the right histogram bin to increment per data point. Rather than doing a binary search on the sorted interval points (which suffers from branch misprediction), we pull in every $32^{nd}$ interval point into a separate subinterval array as a pre-processing step. This is then scanned using SIMD. Once we identify the two sub-interval points between which the data point lies, we scan that specific range of 32-elements in the full interval point array (again using SIMD) and locate the histogram bin to increment. We get overall performance gains of up to 42% (e.g. cosmology dataset) during local kd-tree construction over binary search.

**Choice of bucket size (number of levels):** The global kd-tree has $\log_2 P$ levels, where $P$ is the number of nodes. In the local kd-tree, we create new levels until the number of points in a leaf node is ≤ a predefined bucket size. Once this is reached, we stop creating new levels and pack the points into a bucket. Larger buckets improve construction time, but make querying more expensive (Querying with a bucket is exhaustive). Empirically, we found that a bucket size of 32 gave the best performance.

### B. Distributed KNN querying

The key to efficient nearest neighbor querying is to exploit locality i.e. we need to ensure that points that are nearby geometrically are also localized in memory. This is impossible to do with the exhaustive search approach. We can, however, construct a fully distributed kd-tree that ensures that the geometric domain is partitioned among the nodes. This means that any query only has to go to a small subset of nodes to compute nearest neighbors accurately. This is illustrated in Figure 3. We explain all the steps involved in

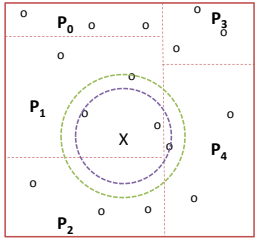

Figure 3. Figure shows the data points (denoted by $o$) in 2D space divided among 5 nodes. Query point is shown as $X$. KNN with $k = 3$ is run in node $P_1$ (owner of $X$). This returns 3 points owned by $P_1$ and a max distance (denoted by green circle around $X$). Only $P_2$ and $P_4$ might own points within this radius. KNN is run on $P_2$ and $P_4$ for $X$ and the closest 3 points are chosen (within purple circle around $X$).

a fully distributed KNN querying algorithm below:

**1) Find owner and send query to that node**: Once a query arrives at any of the nodes, we traverse the global kd-tree to identify the node that owns the domain containing the query (e.g. Node $P_1$ in Figure 3). Since the domain is divided into non-overlapping regions and every node has a copy of the global kd-tree structure, this can be done efficiently. Once the owner is identified, we send the query over to that node.

**2) Find local KNN for queries:** This is the most computationally intensive part of the query process, taking 40-65% of the overall query runtime. We need to traverse the local kd-tree in order to identify the top $k$ closest neighbors among the local points. This traversal is explained in detail in Section III-C. At the end of this stage, we get the local $k$ closest neighbors of $X$. This essentially gives a bound of maximum distance, $r'$ for $X$ that any remote neighbors (owned by other nodes) can not be far from $r'$ as we already have local $k$ neighbors within distance $r'$. This helps pruning the search space, both remote nodes and remote neighbors.

**3) Send query to other nodes within $r'$ of boundary**: We use the $r'$ bound and the global kd-tree to identify which other nodes are within $r'$ distance from the query, $X$. We then send $X$ to those nodes as well.

**4) For received queries, find local KNN and send back**: Every node then checks if it has received any queries from its neighbors as part of the previous step. We perform a local KNN computation for these queries. As we also received $r'$ with each query, local KNN search algorithm (Section III-C) performs early pruning to reduce search space and achieve good performance. The results are then sent back to the owner of the query. In Section V, we show that the local and remote KNN query steps together take 65-85% of overall query runtime.

**5) For received responses, pick the top $k$ among local and remote neighbors**: Once the responses from other nodes are received, we put them all in a heap ordered by the distance from $X$ and pick the top $k$ nearest neighbors.

We perform several optimizations to the process explained earlier. The most important one is batching of queries (computing local KNN of a subset of queries, not all and communicate with remote nodes). This ensures load balance among nodes and better throughput overall. We also perform software pipelining between the stages to facilitate overlap of communication and computation. These optimizations are important for good scaling as the number of nodes increase.

**Algorithm 1** Finding $k$-nearest neighbors from the local kd-tree. Input: kd-tree $T$, Query $q$, $k$, search radius, $r$ (default $r = \infty$). Output: A set, $R$ of $k$ nearest neighbors within $r$.

1: **procedure** FINDKNN($T, q, k, r$)
2:     $r' \leftarrow r$; push $(root, 0)$ into $S$
3:     **while** $S$ is not empty **do**
4:        $(node, d) \leftarrow$ pop from $S$
5:        **if** $node$ is leaf **then**
6:           **for** each particle $x$ in $node$ **do**
7:              compute distance, $d[x]$ of $x$ from $q$
8:              **if** $d[x] < r'$ **then**
9:                  **if** $|H| < k$ **then**
10:                     add $x$ into $H$
11:                     **if** $|H| = k$ **then**
12:                        $r' \leftarrow H.maxi\_dis$
13:                  **else if** $d[x] <$ max distance in $H$ **then**
14:                     replace the topmost point $H$ by $x$
15:                     $r' \leftarrow d[x]$
16:        **else**
17:           **if** $d < r'$ **then**
18:              $d' \leftarrow q[node.dim] - node.median$
19:              $d' \leftarrow \sqrt{d*d + d'*d'}$
20:              $C_1 \leftarrow$ closer child of $node$ from $q$
21:              $C_2 \leftarrow$ other child of $node$
22:              **if** $d' < r'$ **then**
23:                  push $(C_2, d')$ into $S$
24:              push $(C_1, d)$ into $S$
25:     $R \leftarrow H$

### C. Local KNN querying

The algorithm for finding the $k$ nearest neighbors from a local kd-tree, $T$ is given as pseudo-code in Algorithm 1. Finding $k$ nearest neighbors from local kd-trees comes up in two contexts in our application - (1) When a query owner wants to find the top-$k$ neighbors and (2) when nodes that own domains that are adjacent to the query owner check to see if they have closer neighbors than those within the owner node. Both solutions rely on maintaining bounds on the distance to $k^{th}$ neighbor (denoted as $r'$ in Algorithm 1) and progressively refining it while using it to prune regions of the tree to limit exploration. Before traversing a node down in the tree, we always keep track of the minimum distance of the query point to all the points in the node (denoted by $d$ and $d'$ in Algorithm 1) and use this for pruning. Once we reach a leaf node, we find the distance from query to all the points in the bucket. This computation is very SIMD-friendly as the required points are localized in memory as well. In the non-leaf node, we push the closer child, $C_1$ of node from $q$ ($d' < 0$) into stack $S$ later than the other child $C_2$ to get the closer neighbors of $q$ earlier. This essentially helps pruning (Line 17 and 22) the search space for the later processed nodes. We use heap $H$ to keep track of $k$ nearest neighbors found so far.

## IV. EXPERIMENTAL SETUP

We now describe datasets, system, and software tools used to evaluate the performance of different phases in PANDA.

### A. Platforms

Our experiments were performed on Edison, a Cray XC30 supercomputing system, at the National Energy Research Scientific Computing Center (NERSC). The system consists of 5576 compute nodes, each configured with two 12-core Intel® Xeon® [1] E5-2695 v2 processors at 2.4 GHz and 64 GB of 1866-DDR3 memory. Compute nodes communicate using a Cray Aries interconnect that supports injection rates of 10 GB/s bi-directional bandwidth per node. We implemented KNN construction and querying in C/C++ and compiled using Intel® C++ compiler v.15.0.1[2]. The code was parallelized using OpenMP and MPI. We used Intel® MPI library v.5.0.2.

Table I
ATTRIBUTES OF THE COSMOLOGY ($cosmo*$), PLASMA PHYSICS ($plasma*$) AND PARTICLE PHYSICS ($dayabay*$) DATASETS WITH TAKEN TIME BY PANDA IN SECONDS. M, B, C, AND Q STAND FOR MILLION, BILLION, KD-TREE CONSTRUCTION AND QUERYING.

| Name | Particles | Dims | Time (C) | $k$ | Queries (%) | Time (Q) | Cores |
|---|---|---|---|---|---|---|---|
| $cosmo\_small$ | 1.1 B | 3 | 23.3 | 5 | 10 | 12.2 | 96 |
| $cosmo\_medium$ | 8.1 B | 3 | 31.4 | 5 | 10 | 14.7 | 768 |
| $cosmo\_large$ | 68.7 B | 3 | 12.2 | 5 | 10 | 3.8 | 49152 |
| $plasma\_large$ | 188.8 B | 3 | 47.8 | 5 | 10 | 11.6 | 49152 |
| $dayabay\_large$ | 2.7 B | 10 | 4.0 | 5 | 0.5 | 6.8 | 6144 |
| $cosmo\_thin$ | 50 M | 3 | 1.1 | 5 | 10 | 1.1 | 24 |
| $plasma\_thin$ | 37 M | 3 | 1.0 | 5 | 10 | 0.8 | 24 |
| $dayabay\_thin$ | 27 M | 10 | 1.8 | 5 | 0.5 | 3.2 | 24 |

### B. Dataset

*1) Cosmology:* To evaluate the scalability of PANDA, the proposed parallel KNN, we have used datasets produced by three cosmological N-body simulations using Gadget code [14]. We use datasets of three sizes for the large scale runs: referred to as $cosmo\_small$, $cosmo\_medium$, and $cosmo\_large$. The volume of simulations and the number of particles increase synchronously in these three datasets and the number of particles is equal to 1.1, 8.1, and 68.7 billion, respectively. Particle properties in all cosmology datasets include 3D spatial locations and particle velocities. All the datasets have been stored as HDF5 files and each property was stored as one HDF5 1D array dataset. We used spatial locations in our experiments, and the resulting sizes of the datasets are ∼12 GB, ∼96 GB, and ∼0.8 TB, respectively.

*2) Plasma Physics:* The 3D simulation of magnetic reconnection in electron-positron plasma was performed using high-performance fully relativistic PIC code VPIC [15]. The simulation is performed in a 3D domain with open boundary conditions [16] of size $(330 \times 330 \times 132)c/\omega_{pe}$ with $2000 \times 2000 \times 800$ cells. The average initial particle density is 320 particles per species per cell, so that the simulation started tracking roughly two trillion particles (one trillion electrons and one trillion ions) and as it progressed more particles were added due to the open boundaries. Particle properties of interest in this simulation include the 3D spatial locations, kinetic energy $E = m_e c^2(\gamma - 1)$, and individual components of particle velocity. In this study, we have extracted all the data related to particles with $E > 1.1 m_e c^2$ and used spatial locations, thus the resulting size for KNN becomes ∼2.5 TB ($plasma\_large$) that contained 189 billion particles.

*3) Particle physics:* For this study, we analyze data collected from one of the cylindrical antineutrino detectors (Figure 1) at the Daya Bay experiment. The 'records' in the dataset correspond to snapshots in time of a 24 by 8 set of signals recorded from the detector that have undergone calibration but are not reconstructed into derived physics quantities. We encode this data into a 10-dimensional representation using a deep autoencoder neural network with hyperbolic tangent neurons and shape 192-100-10-100-192. We label into 3 classes corresponding to types of physics events that have been previously identified by physicists working on the data. The resulting size of the dataset ($dayabay\_large$) is ∼30 GB with 2.7 billion records.

In our multinode strong scaling and performance analysis, we used all the largest (*_large) dataset. For weak scaling, we used only the cosmology datasets (small, medium, and large using 96 to 6144 cores) as it was the only dataset available that kept the density characteristics similar while increasing the dataset size. For single node experiments, we used relatively smaller datasets (*_thin) with less than $50M$ particles or records from all the three applications.

## V. RESULTS

In this section we present our experimental results for KNN construction and querying, PANDA on Edison using cosmology, plasma physics, and particle physics datasets. We first show results in the multinode and single node settings. We then compare our results with state-of-the-art KNN implementations followed by science results.

---

[1]Intel and Xeon are trademarks of Intel Corporation in the U.S. and/or other countries.

[2]Intel's compilers may or may not optimize to the same degree for non-Intel microprocessors for optimizations that are not unique to Intel microprocessors. These optimizations include SSE2, SSE3, and SSE3 instruction sets and other optimizations. Intel does not guarantee the availability, functionality, or effectiveness of any optimization on microprocessors not manufactured by Intel. Microprocessor-dependent optimizations in this product are intended for use with Intel microprocessors. Certain optimizations not specific to Intel micro-architecture are reserved for Intel microprocessors. Please refer to the applicable product User and Reference Guides for more information regarding the specific instruction sets covered by this notice. Notice revision #20110804

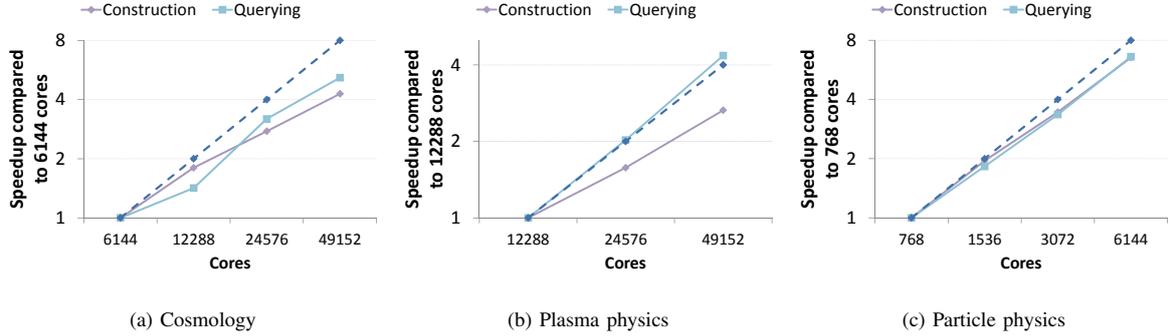

(a) Cosmology  (b) Plasma physics  (c) Particle physics

Figure 4. Strong scaling on $cosmo\_large$ (69B particles), $plasma\_large$ (189B particles) and $dayabay\_large$ (3B records) datasets normalized to the time taken on 6144, 24576, and 768 cores respectively. Dashed lines denote ideal scaling.

*A. Multinode*

*1) Strong Scaling:* We demonstrate the strong scaling performance of kd-tree construction and querying of PANDA using a varying number of cores. We use the largest datasets, $cosmo\_large$ (69B particles), $plasma\_large$ (189B particles), and $dayabay\_large$ (3B records) from the cosmology, plasma physics, and particle physics application domains. Figure 4 shows the results of our strong scaling experiments. Due to memory constraints, we start our experiments at 6144, 12288, and 768 cores for the three datasets respectively. Figures 4(a) shows that the construction and querying phases of PANDA on the $cosmo\_large$ dataset scales by 4.3X and 5.2X as we increase the number of cores from 6144 to 49152 cores (8X). The corresponding scaling number for $plasma\_large$ are 2.7X and 4.4X for construction and querying respectively (Figure 4(b)), when increasing cores from 12288 to 49152 (4X). For the $dayabay\_large$ dataset, the scaling numbers are 6.5X and 6.6X (Figure 4(c)) when increasing cores from 768 to 6144 cores (8X). As we increase the number of cores, we need to partition the dataset more finely. For the construction phase, the depth of the global kd-tree increases as we increase core count leading to additional computation and communication steps. This affects scalability of the construction step (for instance, $plasma\_large$ scales by 2.7X on a 4X increase in core count). In the querying phase, we also see some increase in computation since each query is more likely to be processed by multiple nodes. However, as opposed to construction, querying does not require communication of the original dataset (rather only a small amount of data per query), and hence querying scales better than construction (we see scalability of upto 6.6X vs 5.1X using 8X more cores).

*2) Weak Scaling:* We now show weak scaling results on the cosmology datasets. This is the only scientific dataset for which we have data with varying number of particles with similar density characteristics. We fix the number of particles per node to be ∼250M, and run experiments with 96, 768 and 6144 cores (total of 64X difference in core count). As shown in Figure 5(a), the overall runtime for kd-tree construction and querying increases only by a factor of 2.2X and 1.5X respectively, as we increase the core counts by 64X. Although both construction and querying performance scale well, we observe that querying scales even better than construction. This trend is similar to strong scaling as discussed above.

*3) Runtime breakdown:* Figure 5(b) and 5(c) show the timing breakdown for construction and querying on the largest datasets, $cosmo\_large$, $plasma\_large$ and $dayabay\_large$. We show results on 6144, 12288 and 768 cores respectively, the starting point for our strong scaling experiments. From Figure 5(b), we observe that the global kd-tree construction and particle redistribution steps dominate the overall construction cost, taking over 75% of overall time in the $cosmo\_large$ and $plasma\_large$ datasets. This is due to the fact that these steps require traversals of the entire dataset during median computation and explicit data movement among the nodes. However, in local kd-tree construction, the median computation happens on the already moved local datasets. In addition, it optimizes the communication in the shared memory setting by exploiting the fact that all data is accessible and only moves data indexes rather than the values explicitly in all dimensions. However, for the $dayabay\_large$ dataset which has 10 dimensions, we observe that we spent more time in selecting the right split dimension (as opposed to the 3D $cosmo\_large$ and $plasma\_large$ datasets). This increases local kd-tree computation time, and the global kd-tree construction costs reduces to 58% of overall time. With increasing core counts, the portion of the construction that deals with the global kd-tree construction and redistribution become more dominant. This is due to the fact that while the total height of the global and local sections of the kd-tree remain constant for a fixed data set, increasing core counts means that some levels of the local kd-tree move to the global kd-tree.

To showcase the timing breakdown of querying, we use the same settings as construction (same dataset and core counts). Although the kd-trees was built on the entire datasets, for querying, we use 10%, 10%, and 0.5%

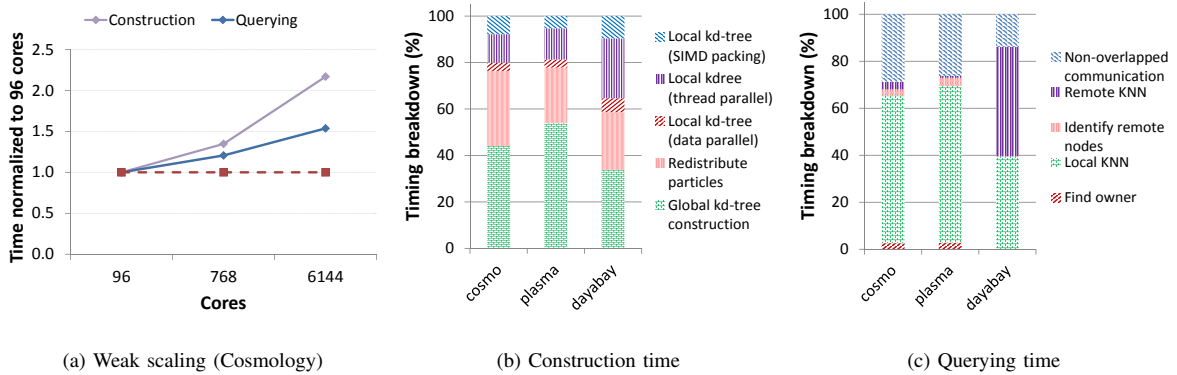

(a) Weak scaling (Cosmology)  (b) Construction time  (c) Querying time

Figure 5. (a) Weak scaling on small, medium and large cosmology datasets using 96, 768, and 6,144 cores, respectively. Dashed line denotes ideal scaling. (b-c) Taken time by different steps of PANDA on the largest cosmology ($cosmo\_large$, 69B particles), plasma physics ($plasma\_large$, 189B particles) and particle physics ($dayabay\_large$, 3B records) using 6144, 12288, and 768 cores, respectively.

random particles of $cosmo\_large$, $plasma\_large$, and $dayabay\_large$ datasets, respectively. As shown in Figure 5(c), local KNN (searching local kd-tree) takes most of the time (up to 67%). This is expected as after reading the queries, each node moves the points to their owners (find owners, taking up to 3% of total time). This was was possible due to the global view of the entire dataset using the global kd-tree. Identifying remote nodes after local KNN, where a query has to be sent, takes an additional 3.5% of the total time. For the $cosmo\_large$ and $plasma\_large$ dataset, we observe that 5% and 9% queries are sent to at least one remote node for computing remote KNN. However, since each query is sent with a radius value (distance to its local $k^{th}$ neighbor), the remote node prunes most of the search space, thus end up taking only up to 3% of the total time by remote KNN. In our implementation, communication overlaps with computation. We measure the non-overlapped communication time and found that to be 29% ($cosmo\_large$) and 26% ($plasma\_large$) of the total time. However, for the $dayabay\_large$ dataset, we observe a different behavior for remote KNN, which takes 46% time. Further investigation shows that even though the kd-tree is balanced, each query ended up with asking an average of 22 remote nodes. This happens due to the fact that a significant number of records are co-located in the particle physics dataset and hence each node ended up searching a huge range, although each query receives only an average of 1 remote nearest neighbor due to heavy pruning technique.

### B. Single Node

*1) Scalability:* We study the scaling of both the construction and querying phases of PANDA on the $cosmo\_thin$, $plasma\_thin$ and $dayabay\_thin$ datasets from the cosmology, plasma physics and particle physics domains respectively. These datasets contain 50M, 37M and 27M particles respectively. We run scaling experiments from 1 to 24 threads on our 24-core CPU, and additionally explore the impact of Simultaneous Multi-Threading (SMT) using a 48-threads run (each core can run 2 threads). Figure 6(a) shows that our kd-tree construction code scales well, achieving 17-20X scaling on 24 cores without using SMT, with a further improvement to 18.3-22.4X scaling on 24 cores using SMT. This near linear scaling shows that we have successfully distributed the construction with varying levels of parallelism to threads with minimal load imbalance. Figure 6(b) shows that the scaling of the querying code is 8.8-12.2X on 24 cores without using SMT. The code is significantly limited by memory accesses, since there is very little work done at each node of the kd-tree traversal. The only computation is the distance computation at the leaf nodes, and this work increases with dimensionality of the dataset ($cosmo\_thin$ and $plasma\_thin$ with 3D data scales worse than $dayabay\_thin$ with 10D data). We notice that there is a significant impact of using SMT on the $cosmo\_thin$ and $plasma\_thin$ datasets (we get additional 1.5-1.7X performance gains), since these datasets have little compute and are highly limited by memory latency. For the $dayabay\_thin$ dataset, the impact of SMT is lower (1.2X performance gain). Overall, we obtain a scaling of 12.9-16.2X scaling on 24 cores. Our performance after using SMT heavily utilizes memory bandwidth (we obtain more than 70% of peak memory bandwidth).

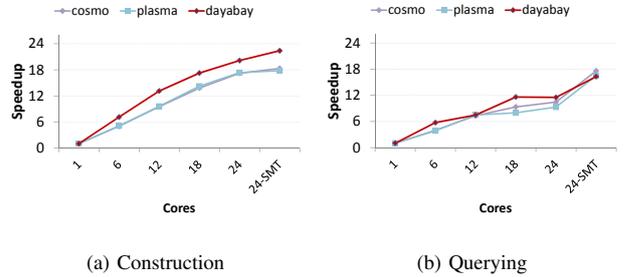

(a) Construction  (b) Querying

Figure 6. Speedup of PANDA construction and querying on single node.

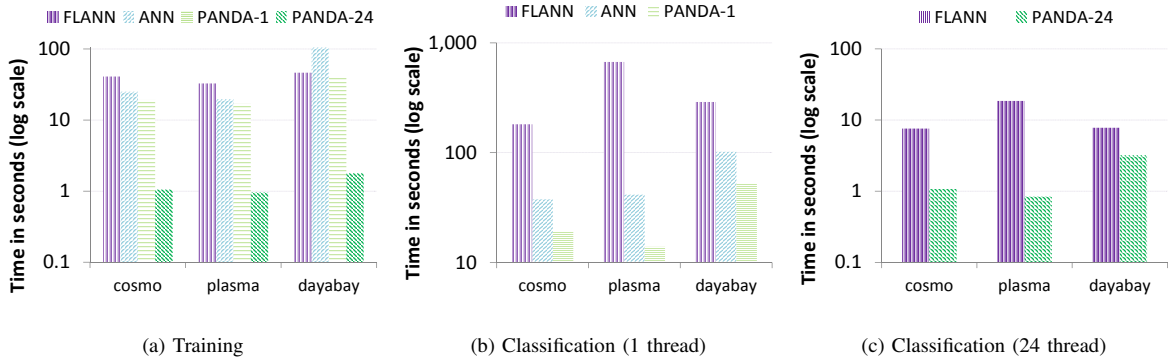

Figure 7. Comparing PANDA with FLANN [13] and ANN [12]. PANDA-1 and PANDA-24 means running on 1 thread and 24 threads respectively

*2) Comparison to Previous Implementations:* We compare the performance of our implementation, PANDA versus the most popular KNN implementations, FLANN [13] and ANN [12]. We use all three *_thin datasets for the comparison. For kd-tree construction, neither FLANN nor ANN can run in parallel. On a single core, we observe that our kd-tree construction code is up 2.2X and 2.6X faster than FLANN and ANN, respectively. FLANN uses variance to select a dimension and then takes an average of the first 100 points over that dimension to compute median during the kd-tree construction. ANN on the other hand uses upper and lower bound of each dimension and select the dimension with maximum difference. Then it takes the average of the lower and upper values of that dimension to compute median. This makes ANN construction faster than FLANN (up to 1.7X except $dayabay$ dataset where the tree becomes more imbalanced and ANN ended up doing more iterations, depth 109 vs. 32 in FLANN). In contrast, our PANDA construction uses sampling based median computation (Section III-A1) leading better balanced trees. We attribute our single core performance improvement of kd-tree construction over FLANN and ANN to implementation optimization including software prefetching, reduced branch misprediction and vectorization in binary search, etc. Using 24 cores, our kd-tree construction becomes more than an order of magnitude faster (39X and 59X) than these two popular implementations. Figure 7(a) shows the performance comparisons of the construction.

We now compare querying on PANDA vs FLANN and ANN. On a single core, we showcase that our implementation is up to 48X and 3X faster than FLANN and ANN, respectively. We observe that although the height of the kd-tree constructed by ANN is higher than FLANN (e.g. on $cosmo\_thin$, the depths are 34 vs 49), querying on FLANN ended up with more node traversals than ANN (e.g. on $cosmo\_thin$, 7X more node traversal). In contrast, our kd-tree has lesser height (21 for $cosmo\_thin$ dataset) and performs less tree node traversals (2X and 12X on $cosmo\_thin$ than FLANN and ANN) leading to more than an order of magnitude performance improvement. Since par-allelizing over queries on shared memory is simple, we use the same outer loop of parallel querying both for FLANN and our querying code. Using 24 cores, we found that our implementation is up to 22X faster than FLANN. Figure 7(b) and 7(c) shows the querying results on 1 core and 24 cores respectively. We have not implemented parallel querying for ANN as the code uses many global variables in different functions making the code unsuitable for parallelization.

### C. Science Results

The primary purpose of this work was to explore the performance of our PANDA implementation on realistic scientific datasets produced by state of the art simulations and experiments. While we have outlined specific classification tasks for astrophysics and plasma physics datasets, we did not have manually labeled classes (or a procedural criteria) on hand for prediction and regression tasks. However, in the case of the Daya Bay dataset, we did have access to 3-class labels annotated by domain science experts. After applying our PANDA system; we observe 87% accuracy in classification performance. To the best of our knowledge, this is the first time machine learning techniques have been used to directly classify an entire raw particle physics dataset without using interim domain specific reconstruction. We can certainly envision more sophisticated classification schemes that utilize spatial weighting of the $k$-neighbors to make even more accurate predictions; but we are quite pleased with the results of this baseline method.

### D. Intel® Xeon Phi™ processor

We now perform multicore and multinode experiments on recently released Intel® Xeon Phi™ processor, commonly known as Knights Landing (KNL). To do so, we performed optimizations such as software prefetching to improve memory performance, taking advantage of SIMD, and adding intrinsics specific to the processor. To compare the performance of our implementations on Intel Xeon Phi processor with the existing manycore state-of-the-art results [17], [18], we took two representative datasets (*psf_mod_mag* and *all_mag*) from [17], [18]. These are photometric data collected from astronomical objects in Sloan Digital Sky Survey [19]. Details on them (top two rows) are given in Table II. Similar to [17], [18], we only experiment KNN querying as the construction dataset is much smaller (taking less than a second) than querying and we used the $k$ value as 10. The machine we used for the experiments consists of 128 KNL nodes, each with 68 cores running at 1.4 GHz.

Table II
DATASETS USED FOR EXPERIMENTS ON INTEL® XEON PHI™ PROCESSOR.

|  | Construction | | Querying | |
| --- | --- | --- | --- | --- |
| Name | Particles | Dims | Particles | Dims |
| $psf\_mod\_mag$ | 2M | 10 | 10M | 10 |
| $all\_mag$ | 2M | 15 | 10M | 15 |
| $cosmo$ | 254M | 3 | 254M | 3 |
| $plasma$ | 250M | 3 | 250M | 3 |

The results of our experiments are given in Figure 8. We first compare our KNN querying implementation on Knights Landing processor (KNL) with the state-of-the-art GPU implementations [17] (Figure 8(a)). Compared to the results reported on an NVIDIA Titan Z GPU, our results on single KNL node are 1.7-3.1X faster. When compared with 4 NVIDIA Titan Z GPU cards, we observe 2.2-3.5X performance boost using 4 KNL nodes. When we compare the scalability on 4 GPU cards vs 4 KNL nodes, we see better scaling on KNL (3.44 vs 3.97 respectively). To demonstrate the scalability of KNL nodes futher (Figure 8(b)), we ran experiments up to 128 KNL nodes achieving almost linear scalability (up to 107X) using the same dataset. Note that we used the same kd-tree on each node (as the kd-tree is small consisting of 2M particles only) similar to the multicard GPU based implementations [17].

However, in scientific computing, kd-trees are larger and unable to fit on a single node. We therefore used the distributed kd-tree implementation (presented above) and perform experiments on separate larger datasets ($cosmo$ and $plasma$) taken from cosmology and plasma physics applications. Details on the datasets can be found in Table II and Section IV-B. Using these datasets, our distributed kd-tree based KNN querying achieved an speedup of 6.6X when we increase the number of KNL nodes from 8 to 64 (8X).

## VI. RELATED WORK

As mentioned earlier, KNN is one of the core machine learning algorithms and therefore there has been a lot of focus on optimizing and parallelizing the algorithm. Although there has been work done on kd-tree based KNN implementations with $O(\log n)$ complexity per query in a shared-memory multi-core setting, most distributed-memory implementations perform a linear time exhaustive search over all points without using kd-trees [9], [10]. [20] presents an approximate distributed KNN algorithm. Even in a shared memory setting, several brute-force approaches have been explored [5], [7]. Most of these linear time algorithms are most suitable for higher dimensional datasets [21], [6] where kd-trees do not work well; and they have only been able to scale up to millions of data points. In contrast, today's bigdata scientific applications come with billions or trillions of particles in a low dimensional space [18].

Recently, there have been some efforts on parallelizing kd-tree construction [22], [23], [24] in the context of dynamic scene ray tracing, where the authors show speedup of up to 7X using 16 cores (while our implementation achieves 17X on 18 cores). Since the authors' codebases are not publicly available, we were unable to compare them directly. A variant of the classical kd-tree (called *buffered kd-tree*) has been proposed recently [18], which parallelizes querying on GPUs given a kd-tree. The idea is to gather queries at leafs of the tree until some number of queries are buffered; these are then processed in parallel. A buffered approach is most useful when latency of queries is not an issue and the total number of queries is very large compared to the data set (e.g. [18] uses scenarios with ∼500X more queries than data points). In scientific applications, the number of queries is likely to be much lesser than the datapoints, and we cannot afford to have large buffers of queries at kd-tree leafs. In contrast, we do not use large buffers and our implementation is up to 3X faster than the buffered approach.

In scientific contexts, both kd-tree construction and query times are important. To the best of our knowledge, no recent work has performed parallel kd-tree construction as well as querying for distributed KNN computation. Since FLANN [13] and ANN [12] are popularly known in the scientific community and their source codes are available online, we compare the performance of both kd-tree construction and querying with our implementation.

## VII. CONCLUSION

In this paper, we present PANDA, a distributed kd-tree based KNN implementation that parallelizes both kd-tree construction and querying on massive datasets of up to 189B particles taken from diverse scientific disciplines. Compared to the state-of-the-art KNN implementations, PANDA is more than an order of magnitude faster on a single node. Further, we show that PANDA scales well up to ∼50,000 cores in both strong and weak scaling senses for all stages of computation. As a result, PANDA can construct kd-trees for 189B particles in ∼48 seconds, and answer 19B KNN queries in ∼12 seconds using 50,000 cores. We believe that scalable and fast kd-tree construction and KNN query times will be critical for the analysis of both scientific simulation and observation datasets. In future, we intend to use PANDA in regression and other scientific applications to gain more insights.

## VIII. ACKNOWLEDGMENTS

This work is supported by the Director, Office of Science, Office of Advanced Scientific Computing Research, of the

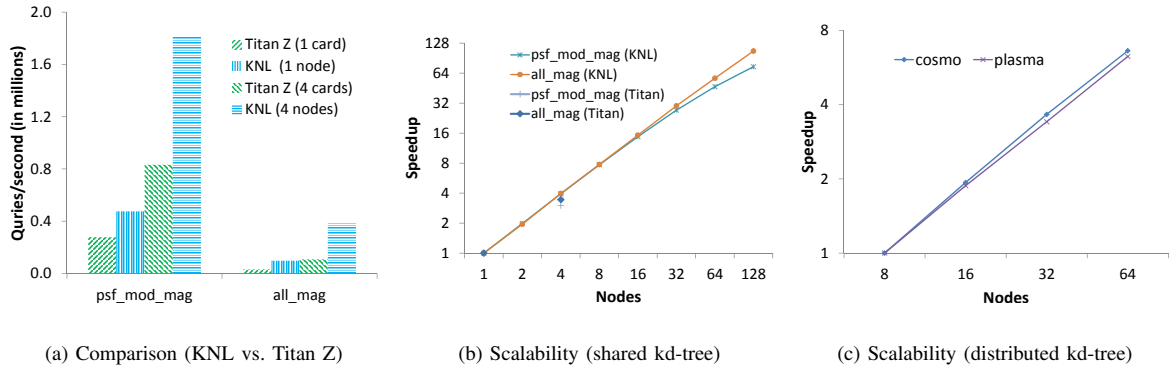

(a) Comparison (KNL vs. Titan Z)  (b) Scalability (shared kd-tree)  (c) Scalability (distributed kd-tree)

Figure 8. Results of KNN querying on Intel® Xeon Phi™ (KNL) processors. (a): Comparing 1 and 4 KNL nodes with 1 and 4 Titan Z cards used in [17], (b) and (c): Strong scaling on KNL nodes using shared and distributed kd-tree respectively

U.S. Department of Energy under Contract No. AC02-05CH11231. This research used resources of the National Energy Research Scientific Computing Center. The authors would like to thank Zarija Lukic and Vadim Roytershteyn for providing access to datasets. We would also like to acknowledge Tina Declerck and Lisa Gerhardt for their assistance in faciliating large scale runs on NERSC platforms.